\documentclass[pra,twocolumn,showpacs,preprintnumbers,showkeys,amsmath,amssymb]{revtex4}
\usepackage{graphicx}   
\usepackage{hyperref}
\usepackage{bm}     
\begin{document}


\title{Quantum jumps and entropy production}

\author{Heinz--Peter Breuer}
\email{breuer@theorie.physik.uni-oldenburg.de}
\affiliation{Fachbereich Physik, Carl von Ossietzky Universit\"at,
         D-26111 Oldenburg, Germany}
\affiliation{Physikalisches Institut, Universit\"at Freiburg,
         D-79104 Freiburg, Germany}
\date{\today}

\begin{abstract}
The irreversible motion of an open quantum system can be
represented through an ensemble of state vectors following a
stochastic dynamics with piecewise deterministic paths. It is
shown that this representation leads to a natural definition of
the rate of quantum entropy production. The entropy production
rate is expressed in terms of the von Neumann entropy and of the
numbers of quantum jumps corresponding to the various decay
channels of the open system. The proof of the positivity and of
the convexity of the entropy production rate is given. Monte Carlo
simulations of the stochastic dynamics of a driven qubit and of a
$\Lambda$-configuration involving a dark state are performed in
order to illustrate the general theory.
\end{abstract}


\maketitle

\section{Introduction}
Dynamical processes in dissipative open systems involve, in
general, irreversible transformations of non-equilibrium states
\cite{GROOT,KEIZER}. An important physical quantity of the theory
of non-equilibrium processes is the entropy production, that is
the rate at which entropy is produced as a result of irreversible
processes \cite{NICOLIS}. In the classical theory the entropy
production provides an important Lyapunov functional of
non-equilibrium stationary states and leads to the principle of
minimal entropy production. It further enables the determination
of non-equilibrium transport properties and Onsager coefficients
(for a recent discussion of the classical theory see,
e.~g.~\cite{DAEMS,BAG}).

In the quantum theory of open systems \cite{TheWork} the entropy
production is usually related to the negative derivative of the
relative entropy with respect to an invariant state. On the ground
of the weak coupling master equation for the open system's density
matrix the corresponding expressions can again be shown to have
useful convexity properties and to satisfy a minimal entropy
production principle \cite{SPOHN}, as well as a kind of quantum
Onsager relations \cite{LENDI}. The concept of relative entropy is
also useful in quantum information theory \cite{NIELSEN}, e.~g. in
the quantification of entanglement \cite{VEDRAL,WERNER} and of the
success of quantum teleportation \cite{BOSE}. A general account of
the theory of quantum entropy may be found in \cite{WERL}.

During the past decade there has been considerable interest in the
application of stochastic wave function methods to the analysis
and simulation of dissipative open quantum systems. By contrast to
the conventional description in terms of the reduced density
matrix, in these methods one considers an ensemble of pure states
whose time-evolution is governed by a stochastic process in the
underlying Hilbert space $\mathcal{H}$ of the open system. In a
particularly interesting method of this kind, developed by several
authors \cite{DALIBARD,DUM,CARMICHAEL}, the wave function
represents a piecewise deterministic process in Hilbert space.
This means that the stochastic evolution consists of smooth,
deterministic parts which are interrupted by instantaneous quantum
jumps. Recently, it has been shown that the stochastic wave
function method can be generalized to enable the treatment of
non-Markovian quantum processes \cite{BKP} and of Bosonic
\cite{CARUSOTTO} and of Fermionic \cite{CHOMAZ} many-body systems
by means of an appropriate stochastic representation in the
doubled Hilbert space $\mathcal{H} \oplus \mathcal{H}$.

The aim of the present paper is to connect the stochastic state
vector dynamics to the production of quantum entropy. To this end,
we apply the entropy balance equation of non-equilibrium
thermodynamics to the von Neumann entropy of the open system. A
quantum jump of the stochastic dynamics is linked to the exchange
of an entropy quantum between system and environment. This enables
one to express the entropy flux in terms of the random processes
which count the numbers of quantum jumps belonging to the various
decay modes of the open system. In this way, a stochastic
representation of the production rate of quantum entropy is
derived from the piecewise deterministic state vector evolution.
It will be demonstrated that the obtained expression for the
entropy production provides a non-negative and convex functional
on the state space of the open system. Furthermore, the stochastic
representation is shown to lead to a Monte Carlo simulation method
for the determination of the quantum entropy production in open
systems.

The paper is organized as follows. In Sec.~II we briefly recall
the dynamical description of open quantum systems in terms of
quantum master equations and of stochastic differential equations.
Sec.~III contains the
construction of the stochastic representation of the entropy
production rate as well as the proof of its positivity and
convexity. Finally, a discussion of the results and examples of
Monte Carlo simulations of the entropy production are given in
Sec.~IV.

\section{Quantum master equations and piecewise deterministic processes}
We consider an open quantum system with reduced density matrix
$\rho=\rho(t)$. The open system is coupled to a heat bath, that is
to an external reservoir in a thermal equilibrium state at
temperature $T$. The free evolution of the system is described by
some Hamiltonian $H_0$. Furthermore, the system may be subjected
to an external time-dependent perturbation, e.~g. an external
radiation field represented by a Hamiltonian $H_p(t)$. Invoking
the Markov approximation, the motion of the reduced system is
assumed to be representable in terms of a quantum master equation
with a generator ${\mathcal{L}}_t$ in Lindblad form
\cite{DAVIES,GORINI,LINDBLAD76} (choosing units such that
$\hbar=1$),
\begin{equation}  \label{LINDBLADeq}
 \frac{d}{d t} \rho = {\mathcal{L}}_t(\rho)
 = -i[H_0+H_p(t),\rho] + {\mathcal{D}}(\rho).
\end{equation}
Since we allow the perturbation to be explicitly time-dependent
the dynamics does not, in general, provide a dynamical semigroup.
However, since ${\mathcal{L}}_t$ is in Lindblad form for each
fixed $t$, the time-evolution defined by Eq.~(\ref{LINDBLADeq})
yields a two-parameter family of quantum dynamical maps
\cite{ALICKI}.

The commutator in Eq.~(\ref{LINDBLADeq}) represents the unitary
part of the motion of the reduced system, while
${\mathcal{D}}(\rho)$ is a super-operator known as the dissipator
of the master equation,
\begin{eqnarray} \label{DISS}
 {\mathcal{D}}(\rho) &=& \sum_i \gamma^-_i
 \left( A^-_i \rho A^+_i
 - \frac{1}{2} \{ A^+_i A^-_i , \rho \} \right)
 \nonumber \\*
 &~& + \sum_i \gamma^+_i
 \left( A^+_i \rho A^-_i
 - \frac{1}{2} \{ A^-_i A^+_i , \rho \} \right).
\end{eqnarray}
The Lindblad operators $A_i^{\pm}$ provide the coupling to the
various decay channels of the open system with corresponding
damping rates $\gamma_i^{\pm}$. We shall assume in the following
that the master equation (\ref{LINDBLADeq}) has been derived from
an underlying microscopic theory through the weak-coupling
approximation. This implies that the Lindblad operators are
obtained as eigen-operators of the unperturbed system Hamiltonian,
\begin{equation} \label{COMMUT}
 [H_0,A_i^{\pm}] = \pm \omega_i A_i^{\pm},
\end{equation}
and satisfy $A_i^+=(A_i^-)^{\dagger}$. The relation (\ref{COMMUT})
also implies that
\begin{equation} \label{COMMUT2}
 [H_0,A_i^+A_i^-] = [H_0,A_i^-A_i^+] = 0.
\end{equation}
The damping rates $\gamma_i^{\pm}$ of the decay modes
are determined through certain reservoir correlation functions 
satisfying the Kubo-Martin-Schwinger condition which leads 
to the relation
\begin{equation}
 \gamma_i^+ = \exp \left( -\omega_i/T \right) \gamma_i^-.
\end{equation}

As mentioned in the introduction the dynamics provided by the
master equation (\ref{LINDBLADeq}) can be represented in terms of
a piecewise deterministic process $\psi(t)$ (PDP) in the Hilbert
space of the open system in the sense that its covariance matrix
\begin{equation} \label{RHODEF}
 \rho(t) \equiv {\mathrm{E}}
 \big[ |\psi(t)\rangle \langle\psi(t)| \big]
 = \int D\psi D\psi^{\ast} P[\psi,t] |\psi\rangle\langle\psi|
\end{equation}
satisfies the master equation (\ref{LINDBLADeq}). Here,
${\mathrm{E}}$ denotes the expectation value of the stochastic
process $\psi(t)$ which is defined in terms of the corresponding
time-dependent probability density functional $P[\psi,t]$.

An appropriate stochastic differential equation for the state
vector dynamics is given by
\begin{eqnarray} \label{PDP-SDGL}
 d\psi(t) &=& -i G_t(\psi(t)) dt \nonumber \\*
 &~& + \sum_i \left( \frac{A^-_i \psi(t)}{||A^-_i\psi(t)||}
 - \psi(t) \right) dN^-_i(t) \nonumber \\*
 &~& + \sum_i \left( \frac{A^+_i \psi(t)}{||A^+_i\psi(t)||}
 - \psi(t) \right) dN^+_i(t).
\end{eqnarray}
The first term on the right-hand side provides the deterministic
evolution periods of the process $\psi(t)$. It corresponds to the
deterministic non-linear Schr\"odinger-type equation
$\dot{\psi}=-iG_t(\psi)$, where
\begin{equation} \label{GDEF}
 G_t(\psi) = \hat{H} \psi
 + \frac{i}{2} \sum_{i} \left( \gamma^-_i
 || A^-_i \psi ||^2 + \gamma^+_i || A^+_i \psi ||^2\right) \psi
\end{equation}
is a non-linear operator with a non-Hermitian Hamiltonian
$\hat{H}=\hat{H}(t)$ given by
\begin{equation} \label{H-HAT}
 \hat{H} = H_0 + H_p(t)
 - \frac{i}{2} \sum_{i} \left( \gamma^-_i A^+_i A^-_i
 + \gamma^+_i A^-_i A^+_i \right).
\end{equation}
The quantities $dN^{\pm}_i(t)$ in Eq.~(\ref{PDP-SDGL}) are the
increments of integer-valued random processes $N^{\pm}_i(t)$
satisfying
\begin{eqnarray}
 dN^k_i(t)dN^l_j(t) &=& \delta_{ij}\delta_{kl}dN^l_j(t),
 \label{POISSON-REL-1} \\
 {\mathrm{E}}\left[ dN^{\pm}_i(t) \right]
 &=& \gamma^{\pm}_i ||A^{\pm}_i\psi(t)||^2 dt,
 \label{POISSON-REL-2}
\end{eqnarray}
where $k,l = \pm$. According to relation (\ref{POISSON-REL-1})
these increments take on the values $0$ or $1$. Moreover, if
$dN^k_i(t)$ is equal to $1$ for a particular pair $(i,k)$, all
other increments are equal to zero. Thus, in view of the
stochastic differential equation (\ref{PDP-SDGL}) the state vector
then undergoes an instantaneous quantum jump of the form
\begin{equation} \label{JUMPS}
 \psi(t) \longrightarrow \frac{A^k_i \psi(t)}{||A^k_i \psi(t)||}.
\end{equation}
The relation (\ref{POISSON-REL-2}) states that the expectation
values of the increments $dN^k_i(t)$ increase linearly with the
time increment $dt$ and are directly proportional to the damping
rates $\gamma_i^k$ and to the square of the norm of $A^k_i
\psi(t)$. Since the latter depends on time through the
deterministic motion, the processes $N_i^k(t)$ are, in general,
inhomogeneous in time. They count the number of quantum jumps of
type $(i,k)$, i.e.\ the number of the jumps (\ref{JUMPS}) with
Lindblad operator $A_i^{k}$.

\section{Definition of the entropy production rate}

\subsection{Entropy flux and numbers of quantum jumps}
An appropriate starting point for the definition of the entropy
production rate $\sigma$ in an open quantum system is provided by
the entropy balance equation,
\begin{equation} \label{BALANCE}
 \sigma = \frac{dS}{dt} + J.
\end{equation}
Here, $S$ is the von Neumann entropy of the open system (the
Boltzmann constant is set equal to $1$),
\begin{equation}
 S = -{\mathrm{tr}} \left\{ \rho \ln \rho \right\}.
\end{equation}
The quantity $J$ represents the entropy flux, that is the entropy
which flows per unit of time from the open system to the
reservoir, in which case we have $J>0$, or from the reservoir to
the open system, such that $J<0$. Thus, according to
Eq.~(\ref{BALANCE}) the quantity $\sigma$ is the rate at which
entropy is produced owing to irreversible processes.

In order to define the entropy production rate $\sigma$ we
therefore need an expression for the entropy flux $J$. To this
end, we recall that the external reservoir causes the wave
function to perform instantaneous quantum jumps given by
(\ref{JUMPS}). According to Eq.~(\ref{COMMUT}) the operator
$A_i^{\pm}$ induces a transition of the system state in which the
energy changes by the amount $\pm\omega_i$. Thus, $A_i^+$
describes upward and $A_i^-$ downward transitions of the system.
Any quantum jump with jump operator $A_i^+$ ($A_i^-$) corresponds
to the absorption (emission) of a quantum $\omega_i$ of energy
from (into) the reservoir. Because the reservoir is assumed to be
in a thermal equilibrium state at temperature $T$, the entropy
exchange $\delta S$ associated with a quantum jump can be defined
by
\begin{equation} \label{ENQUANT}
 \delta S = \frac{\omega_i}{T}.
\end{equation}
To justify this expression one first observes that within the
Markov approximation of the reduced system dynamics the reservoir
correlation functions decay on a time scale which is - by
assumption - short compared to the relaxation time of the open
system. It is therefore reasonable to assume that the energy
exchange in a quantum jump with operator $A_i^{\pm}$ appears as a
change $\mp\omega_i$ of the heat energy of the reservoir. We
further note that the reversible character of the energy exchange
through a quantum jump derives from the fact that for each jump
with operator $A_i^-$ (emission) there also exists the reverse
jump with operator $A_i^+$ (absorption).

Thus we see that in each quantum jump with operator $A_i^{\pm}$ an
{\textit{entropy quantum}} (\ref{ENQUANT}) is absorbed from or
emitted into the reservoir. The integer-valued process
$N_i^{\pm}(t)$ represents the number of quantum jumps with
operator $A_i^{\pm}$ within the time interval $[0,t]$. It follows
that in the time interval $[0,t]$, on the average, the net amount
\begin{equation}
 \sum_i \frac{\omega_i}{T} {\mathrm{E}}
 \left[ N^-_i(t) - N^+_i(t) \right]
\end{equation}
of entropy flows into the reservoir. Consequently, the mean
entropy flux is obtained from the expression
\begin{equation} \label{ENFLUX}
 J = \sum_i \frac{\omega_i}{T} \frac{d}{dt}
 {\mathrm{E}} \left[ N^-_i(t) - N^+_i(t) \right].
\end{equation}
This relation finally leads to the following stochastic expression
of the entropy production rate of the open system,
\begin{equation} \label{SIGMA}
 \sigma = \frac{dS}{dt} + \sum_i \frac{\omega_i}{T} \frac{d}{dt}
 {\mathrm{E}} \left[ N^-_i(t) - N^+_i(t) \right].
\end{equation}
By use of this expression the entropy production rate $\sigma(t)$
can easily be determined by a Monte Carlo simulation of the
stochastic process (\ref{PDP-SDGL}). To this end, one evaluates
the von Neumann entropy by means of Eq.~(\ref{RHODEF}) and records
the numbers $N_i^{\pm}(t)$ of the various quantum jumps carried
out during the time interval $[0,t]$.

\subsection{Positivity of the entropy production}
We note the following important properties of the entropy
production rate $\sigma$ defined by expression (\ref{SIGMA}).
First, it is immediately clear that $\sigma$ is an extensive
quantity: When combining statistically independent systems which
are coupled to the same reservoir, the additivity of the von
Neumann entropy and the additivity of the numbers of quantum jumps
is obvious. Second, Eq.~(\ref{SIGMA}) implies that $\sigma$ does
not depend on the specific representation of the state space of
the reduced system, since neither the von Neumann entropy nor the
number of quantum jumps of a given type changes under unitary
transformations. In particular, the expression (\ref{SIGMA}) is
invariant under a transformation to the interaction picture with
respect to the system Hamiltonian $H_0$. Third, the entropy
production $\sigma$ as defined in this way is always non-negative,
\begin{equation} \label{SECOND-LAW}
 \sigma \geq 0,
\end{equation}
which corresponds to the second law of thermodynamics.

To prove the inequality (\ref{SECOND-LAW}) we first write the
entropy production rate as a functional $\sigma=\sigma[\rho]$ of
the reduced density matrix $\rho$. Invoking the master equation
(\ref{LINDBLADeq}) one finds the time-derivative of the von
Neumann entropy,
\begin{equation} \label{DOTS}
 \frac{dS}{dt} = -{\mathrm{tr}} \left\{ {\mathcal{D}}(\rho)
 \ln \rho \right\}.
\end{equation}
To obtain an appropriate expression for the entropy flux we use
Eqs.~(\ref{ENFLUX}) and (\ref{POISSON-REL-2}), as well as the
representation (\ref{RHODEF}) of the density matrix to get
\begin{equation} \label{ENFLUX-2}
 J = \sum_i \frac{\omega_i}{T} {\mathrm{tr}} \left\{
 \gamma_i^- A_i^+A_i^-\rho - \gamma_i^+ A_i^-A_i^+\rho \right\}.
\end{equation}
On the other hand, employing definition (\ref{DISS}) of the
dissipator and the commutation relations (\ref{COMMUT}) and
(\ref{COMMUT2}) one finds
\begin{eqnarray} \label{ENFLUX-3}
 \lefteqn{ {\mathrm{tr}} \left\{ H_0 {\mathcal{D}}(\rho) \right\}
 \nonumber } \\
 &=& \sum_i \gamma_i^- {\mathrm{tr}} \left\{
 A_i^+ [H_0,A_i^-] \rho + \frac{1}{2}[A_i^+A_i^-,H_0]\rho \right\}
 \nonumber \\*
 &~& + \sum_i \gamma_i^+ {\mathrm{tr}} \left\{
 A_i^- [H_0,A_i^+] \rho + \frac{1}{2}[A_i^-A_i^+,H_0]\rho \right\}
 \nonumber \\*
 &=& -\sum_i \omega_i {\mathrm{tr}} \left\{
 \gamma_i^- A_i^+A_i^-\rho - \gamma_i^+ A_i^-A_i^+\rho \right\}
 \nonumber \\*
 &=& -TJ.
\end{eqnarray}

The commutation relations (\ref{COMMUT}) also imply that the Gibbs
state $\rho_{\mathrm{th}}=\exp (-H_0/T)/Z$ with partition function
$Z$ represents a zero-mode of the dissipator of the master
equation, that is
\begin{equation} \label{ZERO-MODE}
{\mathcal{D}}(\rho_{\mathrm{th}}) = 0.
\end{equation}
Physically, this means that the Gibbs state is a stationary
solution of the master equation without perturbation ($H_p=0$). We
express the Hamiltonian $H_0$ in terms of the Gibbs state,
\begin{equation}
 -H_0/T = \ln \rho_{\mathrm{th}} + \ln Z,
\end{equation}
which, together with Eq.~(\ref{ENFLUX-3}), enables us to write the
entropy flux as follows,
\begin{equation}
 J = {\mathrm{tr}} \left\{ {\mathcal{D}}(\rho) \ln \rho_{\mathrm{th}}
 \right\}.
\end{equation}
Combining this with Eq.~(\ref{DOTS}) we are finally led to the
following expression for the entropy production rate in the state
$\rho$,
\begin{equation} \label{SIG-FUNCTIONAL}
 \sigma[\rho] = -{\mathrm{tr}} \left\{ {\mathcal{D}}(\rho)
 \left( \ln \rho - \ln \rho_{\mathrm{th}} \right) \right\}.
\end{equation}

The inequality (\ref{SECOND-LAW}) can now be shown in two
alternative ways. One way is to apply a theorem by Lindblad
\cite{LINDBLAD75} to the dynamical semigroup
$\Lambda_t=\exp(\mathcal{D}t)$ whose generator is identical to the
dissipator ${\mathcal{D}}(\rho)$ of the master equation, and to
relate the functional (\ref{SIG-FUNCTIONAL}) to the
time-derivative of the relative entropy with respect to the Gibbs
state $\rho_{\mathrm{th}}$. Another possibility is to use Lieb's
theorem \cite{LIEB} to conclude that the map $\rho \mapsto -{\rm
tr} \left\{ {\mathcal{D}}(\rho) \ln \rho \right\}$ is a convex
functional. Since ${\mathrm{tr}}\left\{ {\mathcal{D}}(\rho) \ln
\rho_{\mathrm{th}} \right\}$ is linear in $\rho$, it follows that
the entropy production rate $\sigma=\sigma[\rho]$ represents a
convex functional of the density matrix. With the help of this
property it is easy to demonstrate that the entropy production is
non-negative \cite{SPOHN}. In fact, applying the definition of
convexity to the functional $\rho \mapsto -{\mathrm{tr}} \left\{
{\mathcal{D}}(\rho) \ln \rho \right\}$ one is led to the following
inequality which holds for all $\lambda\in[0,1]$,
\begin{eqnarray*}
 \lefteqn{ -{\mathrm{tr}} \left\{ {\mathcal{D}}
 (\lambda\rho+[1-\lambda]\rho_{\mathrm{th}})
 \ln(\lambda\rho+[1-\lambda]\rho_{\mathrm{th}}) \right\} } \\*
 &~& \leq
 -\lambda{\mathrm{tr}} \left\{ {\mathcal{D}}(\rho)\ln(\rho) \right\}
 -[1-\lambda]{\mathrm{tr}} \left\{ {\mathcal{D}}(\rho_{\mathrm{th}})
 \ln(\rho_{\mathrm{th}}) \right\}.
\end{eqnarray*}
Employing the linearity of ${\mathcal{D}}(\rho)$ and
Eq.~(\ref{ZERO-MODE}) we get
\[
 -\lambda{\mathrm{tr}} \left\{ {\mathcal{D}}(\rho)
 \ln(\lambda\rho+[1-\lambda]\rho_{\mathrm{th}}) \right\}
 \leq -\lambda{\mathrm{tr}} \left\{ {\mathcal{D}}(\rho)\ln(\rho) \right\}.
\]
We divide by $\lambda$ and perform the limit $\lambda
\rightarrow 0$ to arrive at
\[
 -{\mathrm{tr}} \left\{ {\mathcal{D}}(\rho) \ln(\rho_{\mathrm{th}}) \right\}
 \leq -{\mathrm{tr}} \left\{ {\mathcal{D}}(\rho)\ln(\rho) \right\},
\]
which, by virtue of (\ref{SIG-FUNCTIONAL}), is equivalent to
inequality (\ref{SECOND-LAW}).

\section{Discussion and examples}

\subsection{General properties of the entropy production}
 \label{GEN-PROP}
It has been shown in the preceding section that the piecewise
deterministic process (\ref{PDP-SDGL}) for the stochastic state
vector of an open quantum system yields a natural definition of
the entropy production as a measure of the degree of the
irreversibility of the dynamics. In this definition the entropy
flux has been expressed in terms of the random numbers of the
various types of quantum jumps corresponding to the decay channels
of the open system.

The given proof of the positivity of the entropy production rate
demonstrates not only that the entropy production $\sigma$ is
non-negative (see Eq.~(\ref{SECOND-LAW})) but also that it is a
convex functional and that $\sigma(\rho_{\mathrm{th}})=0$ is an
absolute minimum of the entropy production (see
Eq.~(\ref{SIG-FUNCTIONAL})). Hence, in the absence of external
fields ($H_p=0$) the entropy production vanishes when the
equilibrium state is reached, in accordance with the second law of
thermodynamics.

We remark that for vanishing external fields $\sigma$ coincides
with the negative time derivative of the relative entropy with
respect to the Gibbs state $\rho_{\mathrm{th}}$. This is no longer
true when external fields are present. Contrary to the definition
of entropy production in terms of the relative entropy, the above
definition does not assume the existence of a stationary density
matrix.

In the case of a non-vanishing external field ($H_p \neq 0$) one
expects, in general, that the systems stays away from equilibrium
and that, therefore, entropy is continuously produced. An
important physical situation arises (see the examples below) if
there exists a stationary solution $\rho^{\mathrm s}$ of the
master equation in the interaction picture. Since the expression
for the entropy production is not affected by the transformation
to the interaction picture one finds that the entropy production
in the stationary state is given by the constant value
\begin{equation} \label{SIGMA-STAT}
 \sigma^{\mathrm s} = \sum_i \frac{\omega_i}{T} \frac{d}{dt}
 {\mathrm{E}} \left[ N^-_i(t) - N^+_i(t) \right],
\end{equation}
where ${\mathrm{E}}$ denotes the expectation value in the
stationary state. In such a stationary non-equilibrium state
entropy is thus produced at a constant rate $\sigma^{\mathrm s}$.
What happens physically is that the energy supplied to the system
by the external field is ultimately transferred as heat energy to
the reservoir leading to a constant entropy production rate.

It is well-known that the stochastic representation of the master
equation (\ref{LINDBLADeq}) through a stochastic process of the
form (\ref{PDP-SDGL}) is not unique: In general, one can choose,
in an infinite number of different ways, the Lindblad jump
operators and the Hamiltonian operator without changing the form
of the Lindblad generator ${\mathcal{L}}_t$ of the master equation
(\ref{LINDBLADeq}). However, different such choices do change, in
general, the stochastic dynamics (\ref{PDP-SDGL}). This fact is
well understood in the framework of continuous measurement theory
according to which different choices for the jump operators
correspond to different measurement schemes used to monitor the
open system \cite{WISEMAN1,WISEMAN2}. Our construction of
expression (\ref{SIGMA}) for the entropy production rate
presupposes, however, that the jump operators $A_i^{\pm}$ are
eigen-operators of the system Hamiltonian, satisfying the
commutation relation (\ref{COMMUT}). It is this property which
enables one to establish the connection between the stochastic
expression (\ref{SIGMA}) and the functional
(\ref{SIG-FUNCTIONAL}). Physically, it is the eigen-operator
property which allows one to associate a certain quantum jump of
type $(i,k)$ with the exchange of a quantum $\omega_i$ of energy
with the thermal reservoir.

In the case of a non-degenerate frequency spectrum $\{\omega_i\}$
the requirement provided by the commutation relations
(\ref{COMMUT}) uniquely fixes the jump operators and, thus, the
stochastic state vector dynamics. If a certain frequency
$\omega_i$ is degenerate one has the freedom to replace the
corresponding jump operators $A_{i\alpha}^{\pm}$ (the index
$\alpha$ labels different Lindblad operators belonging to the same
frequency $\omega_i$) by new operators
\begin{equation} \label{MIXING}
 \tilde{A}_{i\alpha}^{-}
 = \sum_{\beta} u_{\alpha\beta} A_{i\beta}^{-}, \;\;\;
 \tilde{A}_{i\alpha}^{+}
 = \left( \tilde{A}_{i\alpha}^{-} \right)^{\dagger},
\end{equation}
where $u_{\alpha\beta}$ is a unitary matrix. However, in view of
Eq.~(\ref{ENFLUX-2}) it is immediately clear that expression
(\ref{SIGMA}) remains invariant under this replacement, showing
the uniqueness of the stochastic expression of the entropy
production rate.

\subsection{Entropy production rate of a driven qubit}
We conclude the paper by two examples which illustrate several of
the features discussed above. The first example is two-state
system \cite{GARDINER}, a qubit consisting of an excited state
$|e\rangle$ and a ground state $|g\rangle$ with corresponding
energies $E_e$ and $E_g$. The qubit is coupled to a thermal
reservoir and to a single-mode driving field on resonance with the
transition frequency $\omega=E_e-E_g$. The following discussion
will be carried out in the interaction picture. The external field
is treated in the rotating wave approximation such that the
corresponding interaction Hamiltoninan $H_p$ becomes
time-independent in the interaction picture.

The stochastic dynamics given by Eq.~(\ref{PDP-SDGL}) is defined
here through two jump operators, given by the raising and lowering
operators
\begin{equation}
 A^+ = |e\rangle\langle g|, \qquad
 A^- = |g\rangle\langle e|,
\end{equation}
with stochastic processes $N^{\pm}(t)$ for the numbers of the
corresponding jumps. The non-Hermitian Hamiltonian (\ref{H-HAT})
therefore takes the form
\begin{equation}
 \hat{H} = -\frac{\Omega}{2} \left( A^+ + A^- \right)
 -\frac{i}{2}
 \left(\gamma^-A^+A^-+\gamma^+A^-A^+\right),
\end{equation}
where the Rabi frequency $\Omega$ is assumed to be real. Equation
(\ref{SIGMA-STAT}) then yields the following entropy production
rate in the stationary state,
\begin{equation} \label{STAT-EN}
 \sigma^{\mathrm s} = -\frac{\omega}{T} \frac{\Omega^2}{\gamma} S_3
 =\frac{\omega}{T} \frac{\gamma^{-}-\gamma^{+}}{2+(\gamma/\Omega)^2},
\end{equation}
where $\gamma=\gamma^{-}+\gamma^{+}$. Obviously, $\sigma^{\mathrm
s}$ is proportional to the inversion $S_3=\rho^{\mathrm
s}_{ee}-\rho^{\mathrm s}_{gg}$ in the stationary state.

\begin{figure}[htb]
\includegraphics[width=\linewidth]{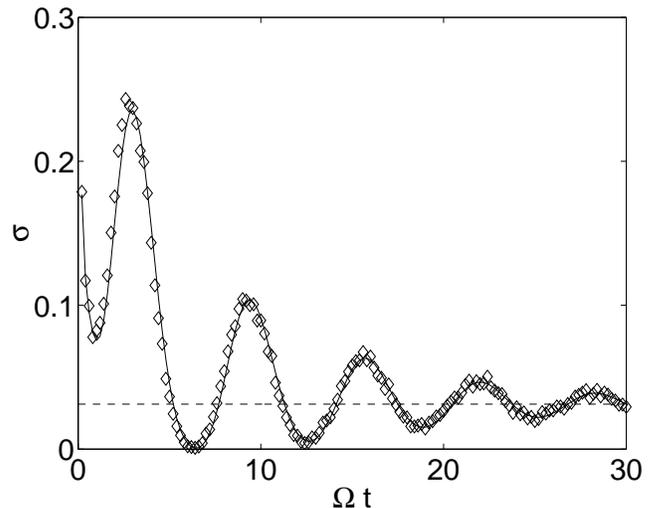}
\caption{\label{figure1}The entropy production rate $\sigma(t)$ of
a driven qubit. Diamonds: Results of a Monte Carlo simulation of
the stochastic differential equation (\ref{PDP-SDGL}) using a
sample of $10^5$ realizations with parameters
$\gamma^-/\Omega=0.1$ and $\omega/T=1.0$. Continuous line: The
entropy production rate obtained from the solution of the master
equation (\ref{LINDBLADeq}). Dashed line: The stationary rate
$\sigma^{\mathrm s}$ according to Eq.~(\ref{STAT-EN}).}
\end{figure}

Figure \ref{figure1} shows the result of a Monte Carlo simulation
of the PDP defined by Eq.~(\ref{PDP-SDGL}) and compares it with
the solution of the density matrix equation (\ref{LINDBLADeq}). In
the stochastic simulation the entropy production rate has been
determined with the help of Eq.~(\ref{SIGMA}), where both the von
Neumann entropy $S$ and the expectation values of the numbers
$N^{\pm}(t)$ of quantum jumps have been estimated from a sample of
$10^5$ realizations of the PDP. The initial condition has been
chosen to be the ground state $|g\rangle$ of the qubit. In the
absence of the driving field the entropy production decreases to
zero during the approach to the thermal equilibrium state.
However, as can be seen from the figure, in the presence of an
external driving the entropy production oscillates and converges
to the stationary value given by Eq.~(\ref{STAT-EN}).

\subsection{$\Lambda$-configuration with a dark state}
As our second example we investigate a two-level atom with
transition frequency $\omega$. Both levels are threefold
degenerate forming manifolds with total angular momentum $J_e=1$
(excited level) and $J_g =1$ (ground level). We introduce the
energy eigenstates $|g,m_g \rangle$ and $|e,m_e \rangle$ which are
simultaneously eigenstates of the $z$-component of the atomic
angular momentum operator with eigenvalues $m_g$ and $m_e$,
respectively. In addition to the coupling to a thermal reservoir
the atom is subjected to a resonant driving field which is
linearly polarized in the $y$-direction. If we take some initial
state in the manifold spanned by the states $|g,m_g=\pm 1\rangle$
we then find that the dynamics is confined to the subspace spanned
by the basis states
\begin{equation} \label{LAMBDA-BASIS}
 |e,m_e=0\rangle, \qquad |g,m_g=+1\rangle, \qquad
 |g,m_g=-1\rangle,
\end{equation}
forming a level scheme known as $\Lambda$-configuration
\cite{SCULLY}.

\begin{figure}[htb]
\includegraphics[width=\linewidth]{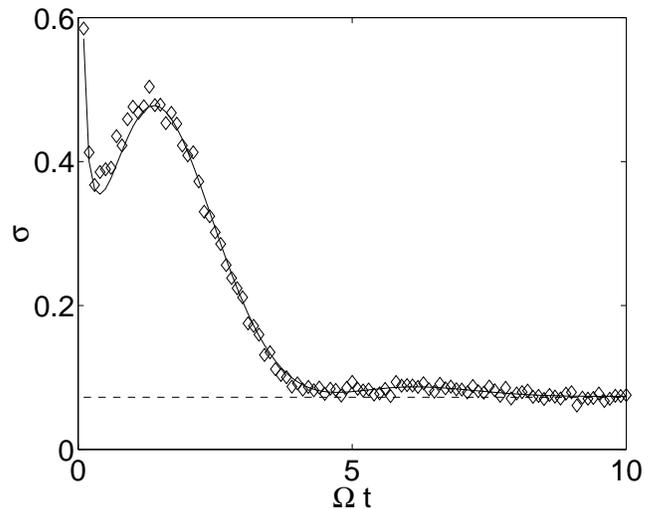}
\caption{\label{figure2}The entropy production rate $\sigma(t)$ of
a driven $\Lambda$-configuration involving a dark state. Diamonds:
Monte Carlo simulation of the corresponding stochastic
differential equation (\ref{PDP-SDGL}) with $10^5$ realizations
and parameters $\gamma^-/\Omega=1$ and $\omega/T=1.25$. Continuous
line: The entropy production rate obtained from the solution of
the master equation. Dashed line: The stationary entropy
production rate $\sigma^{\mathrm s}$ given in
Eq.~(\ref{LAMBDA-SIGMA-STAT}).}
\end{figure}

The dynamics of the system will again be studied in the
interaction picture. It can be described by the four jump
operators,
\begin{equation} \label{LAMBDA-JUMP-OPS}
 A^-_1 = \frac{1}{\sqrt{2}} |g,-1\rangle\langle e,0|, \qquad
 A^-_2 = \frac{1}{\sqrt{2}} |g,+1\rangle\langle e,0|,
\end{equation}
and $A_1^+=(A^-_1)^{\dagger}$, $A_2^+=(A^-_2)^{\dagger}$. The
corresponding matrix representation of the non-Hermitian
Hamiltonian (\ref{H-HAT}) in the basis (\ref{LAMBDA-BASIS}) reads
\begin{equation} \label{LAMBDA-H-HAT}
 \hat{H} = -\frac{i}{2}
 \left( \begin{array} {ccc}
 \gamma^- & \Omega & -\Omega \\
 -\Omega & \frac{\gamma^+}{2} & 0  \\
 \Omega & 0 & \frac{\gamma^+}{2}
 \end{array} \right).
\end{equation}
A numerical simulation of the dynamics is shown in
Fig.~\ref{figure2}. In the Monte Carlo simulation of the
corresponding stochastic differential equation (\ref{PDP-SDGL})
the quantity $\sigma(t)$ has been determined with the help of
Eq.~(\ref{SIGMA}) by recording the stochastic numbers
$N_{1,2}^{\pm}(t)$ of upward and downward transitions in a sample
of realizations of the process.

The stationary entropy production rate may be found from
Eq.~(\ref{SIGMA-STAT}) with the result
\begin{equation} \label{LAMBDA-SIGMA-STAT}
 \sigma^{\mathrm s} = \frac{\omega}{T} \gamma^+ \rho_{23}^s
 = \frac{\omega}{T} \gamma^+
 \frac{\gamma^- - \gamma^+}{2\gamma^- + 4\gamma^+ +
 \frac{\gamma^+}{\Omega^2}(\gamma^-+\frac{\gamma^+}{2})^2}.
\end{equation}
This equation shows that $\sigma^{\mathrm s}$ is proportional to
the matrix element $\rho_{23}^s=\langle g,+1 |\rho^{\mathrm
s}|g,-1\rangle$ of the stationary density matrix $\rho^{\mathrm
s}$. We observe that, similarly to the case of the driven qubit
(see Eq.~(\ref{STAT-EN})), $\sigma^{\mathrm s}$ approaches a
finite value as $\Omega\longrightarrow\infty$. For zero Rabi
frequency, $\Omega=0$, the stationary entropy production rate
vanishes, in accordance with our general discussion. Moreover, the
entropy production vanishes for infinite temperatures for both the
driven qubit and the $\Lambda$-configuration.

By contrast to the driven qubit, however, in the case of the
$\Lambda$-configuration the limit $T\longrightarrow 0$,
corresponding to a reservoir at zero temperature, leads to a
vanishing entropy production rate. In fact, in this limit we have
$\gamma^- \longrightarrow \gamma_0$, $\rho_{23}^s\longrightarrow
1/2$ and, therefore,
\begin{equation} \label{LIMIT-DS}
 \lim_{T\rightarrow 0} \sigma^{\mathrm s}
 = \lim_{T\rightarrow 0} \frac{\omega}{T} e^{-\omega/T}
 \gamma_0 \frac{1}{2} = 0.
\end{equation}
This behaviour can easily be understood if we note that the matrix
element $\rho_{23}$ of the density matrix describes the coherence
of the state
\begin{equation} \label{DARK-STATE}
 |\psi_{\mathrm d}\rangle =
 \frac{1}{\sqrt{2}} \left( |g,+1\rangle + |g,-1\rangle \right).
\end{equation}
This state is known as dark state since it does not couple to the
driving field on account of a quantum interference effect.

According to Eq.~(\ref{LAMBDA-SIGMA-STAT}) the stationary entropy
production rate is thus proportional to the coherence of the dark
state in the stationary state $\rho^{\mathrm s}$. At zero
temperature, any initial state is driven into the dark state under
the dynamics. The reason of this fact is that $|\psi_{\mathrm
d}\rangle$ represents a zero-mode of the non-Hermitian Hamiltonian
(\ref{LAMBDA-H-HAT}) for $T=0$, that is for $\gamma^+=0$.
Equation~(\ref{LIMIT-DS}) therefore states that the entropy
production rate vanishes in the dark state. This follows
immediately from the fact that $|\psi_{\mathrm d}\rangle$ belongs
to the zero temperature ground state manifold.

Finally, we remark that this system provides an example of a
system with a degenerate frequency spectrum, mentioned at the end
of Sec.~\ref{GEN-PROP}. Both jump operators given in
(\ref{LAMBDA-JUMP-OPS}) belong to the same transition frequency
$\omega$. In a continuous measurement interpretation they
correspond to a measurement of the angular momentum of the emitted
quanta along the $z$-direction \cite{MOLMER}. The measurement
along any other direction amounts to a transformation of these
jump operators which is of the form (\ref{MIXING}) and which does
not change the expression for the entropy production rate. The
latter is therefore seen to be independent of the measurement
scheme.


\begin{thebibliography}{99}
\bibitem{GROOT} S.~R. de Groot and P.~Mazur,
                \textit{Non-Equilibrium Thermodynamics},
                (North-Holland, Amsterdam, 1962).
\bibitem{KEIZER} J.~Keizer,
                 \textit{Statistical Thermodynamics of Nonequilibrium
                 Processes}, (Springer-Verlag, New York, 1987).
\bibitem{NICOLIS} G.~Nicolis and I.~Prigogine,
                  \textit{Self-Organization in Nonequilibrium Systems},
                  (Wiley, New York, 1977).
\bibitem{DAEMS} D.~Daems and G.~Nicolis, Phys. Rev. E \textbf{59}, 4000 (1999).
\bibitem{BAG} Bidhan Chandra Bag, Phys. Rev. E \textbf{66}, 026122 (2002).
\bibitem{TheWork} H.~P.~Breuer and F.~Petruccione,
                  \textit{The Theory of Open Quantum Systems},
                  (Oxford University Press, Oxford, 2002).
\bibitem{SPOHN} H.~Spohn, J. Math. Phys. \textbf{19}, 1227 (1978).
\bibitem{LENDI} K.~Lendi, Phys. Rev. A {\textbf{34}}, 662 (1986).
\bibitem{NIELSEN} M.~A. Nielsen and I.~L. Chuang,
                  \textit{Quantum Computation and Quantum Information},
                  (Cambridge University Press, Cambridge, 2000).
\bibitem{VEDRAL} L.~Henderson and V.~Vedral, Phys. Rev. Lett. \textbf{84},
                 2263 (2000).
\bibitem{WERNER} K.~Audenaert, B.~De Moor, K.~G.~H.~Vollbrecht,
                 and R.~F.~Werner, Phys. Rev. A \textbf{66}, 032310 (2002).
\bibitem{BOSE} G.~Bowen and S.~Bose, Phys. Rev. Lett. \textbf{87},
               267901 (2001).
\bibitem{WERL} A.~Werl, Rev. Mod. Phys. \textbf{50}, 221 (1978).
\bibitem{DALIBARD} J.~Dalibard, Y.~Castin, and K.~M{\o}lmer,
                   Phys. Rev. Lett. \textbf{68}, 580 (1992).
\bibitem{DUM} R.~Dum, P.~Zoller, and H.~Ritsch,
              Phys. Rev. A \textbf{45}, 4879 (1992).
\bibitem{CARMICHAEL} H. Carmichael, \textit{An Open Systems Approach
                     to Quantum Optics}, Lecture Notes in Physics m18
                     (Springer-Verlag, Berlin, 1993).
\bibitem{BKP} H.~P.~Breuer, B.~Kappler, and F.~Petruccione,
              Phys. Rev. A \textbf{59}, 1633 (1999).
\bibitem{CARUSOTTO} I.~Carusotto, Y.~Castin, and J.~Dalibard,
                    Phys. Rev. \textbf{63}, 023606 (2001).
\bibitem{CHOMAZ} O.~Juillet and Ph.~Chomaz,
                 Phys. Rev. Lett. \textbf{88}, 142503 (2002).
\bibitem{DAVIES} E.~B.~Davies,
                 Commun. Math. Phys. \textbf{39}, 91 (1974).
\bibitem{GORINI} V.~Gorini, A.~Kossakowski, and E.~C.~G.~Sudarshan,
                 J. Math. Phys. \textbf{17}, 821 (1976).
\bibitem{LINDBLAD76} G.~Lindblad, Commun. Math. Phys. \textbf{48}, 119 (1976).
\bibitem{ALICKI} R.~Alicki and K.~Lendi,
                 \textit{Quantum Dynamical Semigroups and Applications},
                 Volume 286 of {\em Lecture Notes in Physics},
                 (Springer-Verlag, Berlin, 1987).
\bibitem{LINDBLAD75} G.~Lindblad, Commun. Math. Phys. \textbf{40}, 147 (1975).
\bibitem{LIEB} E.~H. Lieb, J. Math. Phys. \textbf{14}, 1938 (1973).
\bibitem{WISEMAN1} H.~M.~Wiseman and G.~J.~Milburn,
                   Phys. Rev. A \textbf{47}, 642 (1993).
\bibitem{WISEMAN2} H.~M.~Wiseman and G.~J.~Milburn,
                   Phys. Rev. A \textbf{47}, 1652 (1993).
\bibitem{GARDINER} C.~W.~Gardiner and P.~Zoller, \textit{Quantum Noise},
                   second edition, (Springer-Verlag, Berlin, 2000).
\bibitem{SCULLY} M.~O.~Scully and M.~S.~Zubairy, \textit{Quantum Optics},
                 (Cambridge University Press, Cambridge, 1997).
\bibitem{MOLMER} K.~M{\o}lmer, Y.~Castin, and J.~Dalibard,
                 J. Opt. Soc. Am. B \textbf{10}, 524 (1993).
\end{thebibliography}
\end{document}